\documentclass[sigconf]{acmart}
\AtBeginDocument{%
  }

\copyrightyear{2026}
\acmYear{2026}
\setcopyright{cc}
\setcctype{by}
\acmConference[ICSE-NIER '26]{2026 IEEE/ACM 48th International Conference on Software Engineering}{April 12--18, 2026}{Rio de Janeiro, Brazil}
\acmBooktitle{2026 IEEE/ACM 48th International Conference on Software Engineering (ICSE-NIER '26), April 12--18, 2026, Rio de Janeiro, Brazil}
\acmPrice{}
\acmDOI{10.1145/3786582.3786803}
\acmISBN{979-8-4007-2425-1/2026/04}

\usepackage{multirow}
\usepackage{makecell}
\usepackage{subfigure}
\usepackage{enumitem}
\usepackage{balance}
\usepackage{listings}
\usepackage{color}
\usepackage{algorithm}

\usepackage{amssymb}

\begin{document}
	
\author{Lingzhe Zhang}
\affiliation{%
	\institution{Peking University}
	\city{Beijing}
	\country{China}}
\orcid{0009-0005-9500-4489}
\email{zhang.lingzhe@stu.pku.edu.cn}

\author{Tong Jia$^{\ast}$}
\thanks{$^{\ast}$Corresponding author}
\affiliation{%
	\institution{Peking University}
	\city{Beijing}
	\country{China}}
\orcid{0000-0002-5946-9829}
\email{jia.tong@pku.edu.cn}

\author{Yunpeng Zhai}
\affiliation{%
	\institution{Alibaba Group}
	\city{Beijing}
	\country{China}}
\orcid{0000-0002-3344-4543}
\email{zhaiyunpeng.zyp@alibaba-inc.com}

\author{Leyi Pan}
\affiliation{%
	\institution{Tsinghua University}
	\city{Beijing}
	\country{China}}
\orcid{0009-0008-0859-2203}
\email{panly24@mails.tsinghua.edu.cn}

\author{Chiming Duan}
\affiliation{%
	\institution{Peking University}
	\city{Beijing}
	\country{China}}
\orcid{0009-0008-4422-6323}
\email{duanchiming@stu.pku.edu.cn}

\author{Minghua He}
\affiliation{%
	\institution{Peking University}
	\city{Beijing}
	\country{China}}
\orcid{0000-0003-4439-9810}
\email{hemh2120@stu.pku.edu.cn}

\author{Pei Xiao}
\affiliation{%
	\institution{Peking University}
	\city{Beijing}
	\country{China}}
\orcid{0000-0002-1674-0308}
\email{xiaopei@stu.pku.edu.cn}

\author{Ying Li$^{\ast}$}
\affiliation{%
	\institution{Peking University}
	\city{Beijing}
	\country{China}}
\orcid{0000-0002-6278-2357}
\email{li.ying@pku.edu.cn}

\renewcommand{\shortauthors}{Lingzhe Zhang, et al.}

\title{Hypothesize-Then-Verify: Speculative Root Cause Analysis for Microservices with Pathwise Parallelism}

\begin{abstract}
Microservice systems have become the backbone of cloud-native enterprise applications due to their resource elasticity, loosely coupled architecture, and lightweight deployment. Yet, the intrinsic complexity and dynamic runtime interactions of such systems inevitably give rise to anomalies. Ensuring system reliability therefore hinges on effective root cause analysis (RCA), which entails not only localizing the source of anomalies but also characterizing the underlying failures in a timely and interpretable manner. Recent advances in intelligent RCA techniques, particularly those powered by large language models (LLMs), have demonstrated promising capabilities, as LLMs reduce reliance on handcrafted features while offering cross-platform adaptability, task generalization, and flexibility. However, existing LLM-based methods still suffer from two critical limitations: (a) limited exploration diversity, which undermines accuracy, and (b) heavy dependence on large-scale LLMs, which results in slow inference. To overcome these challenges, we propose SpecRCA, a speculative root cause analysis framework for microservices that adopts a \textit{hypothesize-then-verify} paradigm. SpecRCA first leverages a hypothesis drafting module to rapidly generate candidate root causes, and then employs a parallel root cause verifier to efficiently validate them. Preliminary experiments on the AIOps 2022 dataset demonstrate that SpecRCA achieves superior accuracy and efficiency compared to existing approaches, highlighting its potential as a practical solution for scalable and interpretable RCA in complex microservice environments.
\end{abstract}



\maketitle

\section{Introduction}

Microservice systems have become increasingly prevalent in cloud-native enterprise applications due to their scalability, modularity, and lightweight deployment~\cite{zhang2025survey}. Yet their complex, highly interconnected runtime behavior makes anomalies unavoidable, where a single failure may rapidly cascade across dependent services and trigger system-wide disruption~\cite{mendoncca2019developing, waseem2021design, zhang2024time}. Ensuring reliability therefore demands effective root cause analysis (RCA) — identifying the origin of anomalies and characterizing failure mechanisms in a timely and interpretable manner.

Manually troubleshooting root causes heavily rely on expert knowledge, which is not only inefficient but also becomes increasingly inadequate as system scale expands and business scenarios diversify~\cite{wang2023interdependent, zhang2024failure, yu2024survey, sun2025interpretable, zhu2024hemirca, xie2024microservice}. Consequently, achieving intelligent and automated diagnosis has become a core requirement for ensuring the stability and reliability of software systems. Existing diagnostic methods can be broadly categorized into rule-based approaches, machine learning methods with feature engineering, and deep learning models. Rule-based approaches depend on manually crafted rules, which are difficult to maintain in highly dynamic environments~\cite{xu2017logdc, lu2017log}. Machine learning methods improve the degree of automation to some extent, but remain highly dependent on domain expertise, with feature extraction and construction being both labor-intensive and costly~\cite{lin2016log, zhang2021onion, amar2019mining, zhang2024reducing}. Deep learning methods mitigate the reliance on manual feature design, but often suffer from poor interpretability and limited transferability across platforms and tasks~\cite{zhang2024multivariate, li2022swisslog, sui2023logkg, zhang2024towards, zhang2025log, zhang2025microremed}. Overall, these methods face common challenges, including lack of cross-platform generality, insufficient adaptability to dynamic system states, and inadequate automation to meet the demands of large-scale system operations.

Recently, the rapid advancement of LLMs has introduced new opportunities for failure diagnosis. With their powerful knowledge representation and reasoning capabilities, LLMs can reduce dependence on handcrafted features while offering strong cross-platform adaptability, task generalization, and flexibility, thereby alleviating the limitations of conventional methods~\cite{zhang2024mabc, wang2024rcagent, pei2025flow, li2025coca, wang2025tamo, ren2025multi, roy2024exploring, shi2024enhancing, xie2024cloud, han2024potential, zhang2025adaptive, zhang2025thinkfl, zhang2025scalalog, zhang2025agentfm, zhang2024automated, shan2024face, zhai2025agentevolver}. Existing efforts can be broadly grouped into three categories: (i) early explorations that demonstrate the potential of LLMs for root cause analysis~\cite{sarda2024leveraging, roy2024exploring, shi2024enhancing, xie2024cloud, han2024potential, zhang2025agentfm, zhang2025xraglog}, (ii) systematic solutions that adopt multi-agent architectures~\cite{zhang2024mabc, wang2024rcagent, pei2025flow, wang2025tamo, ren2025multi, zhang2025adaptive, zhang2025thinkfl}, and (iii) knowledge-enhanced approaches such as retrieval-augmented generation (RAG)~\cite{zhang2025scalalog, zhang2024automated, sarda2024leveraging, shan2024face, li2025coca}. Despite these promising directions, the practical adoption of LLM-based RCA in real-world microservices still faces critical challenges:

\begin{figure*}[htbp]
	\centering
	\includegraphics[width=1\linewidth]{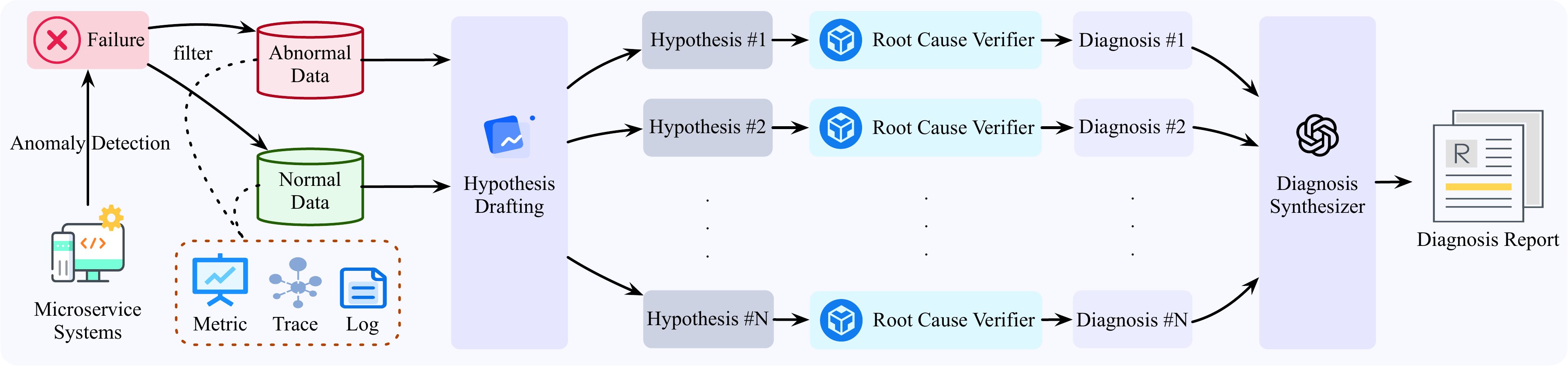}
	\caption{The overview of SpecRCA. SpecRCA follows a Hypothesize-Then-Verify paradigm: it first employs a Hypothesis Drafting module to quickly generate potential hypotheses, and then uses a Root Cause Verifier to validate them in parallel.}
	\label{fig: architecture}
	\vspace{-1.0em}
\end{figure*}

\begin{itemize}[leftmargin=*]
	\item \textbf{Limited Exploration Diversity Leading to Suboptimal Accuracy.} Existing LLM-based RCA methods — including multi-agent voting (e.g., mABC~\cite{zhang2024mabc}) and training-time exploration enhancement (e.g., ThinkFL~\cite{zhang2025thinkfl}) — still tend to converge on similar reasoning paths during inference. As a result, alternative hypotheses remain underexplored, leading to accuracy ceilings even when additional agents or repeated evaluations are used.
	\item \textbf{Overreliance on Large-Scale LLMs Resulting in Slow Inference.} Recent approaches remain dependent on large closed-source or high-parameter LLMs (e.g., Claude, GPT-4o) to perform knowledge reasoning, long-context processing, or multi-round interaction (e.g., RCLAgent~\cite{zhang2025adaptive}, COCA~\cite{li2025coca}). Even when using mid-sized models (e.g., QwQ-32B), multi-agent communication introduces multi-turn latency, making current solutions impractically slow for real-time microservice RCA.
\end{itemize}

To address these challenges, we propose \textbf{SpecRCA}, a speculative root cause analysis framework for microservices that incorporates \emph{pathwise parallelism} and a \emph{hypothesize-then-verify} paradigm. 

For the first challenge, SpecRCA employs a \textit{Hypothesis Drafting Model} that does not aim to produce the most accurate root cause list, but instead enumerates a broad and inclusive set of potential hypotheses. Each candidate root cause is explicitly treated as a hypothesis. For every hypothesis, a \textit{Root Cause Verifier} independently evaluates its plausibility and provides supporting reasoning. Finally, a \textit{Diagnosis Synthesizer} integrates all verification outcomes into a final diagnosis report. This enforced evaluation of each hypothesis ensures substantial exploration diversity, mitigating the tendency of prior methods to converge prematurely on shallow or homogeneous reasoning paths. 

For the second challenge, SpecRCA strategically leverages models of different sizes to balance efficiency and accuracy. The Hypothesis Drafting Model is implemented as a lightweight machine learning model that enumerates possible root causes through data-driven analysis. The hypothesis verification stage employs fine-tuned lightweight LLMs (parameters $\textless$ 3B), which execute verification tasks in parallel to accelerate inference. Only the Diagnosis Synthesizer relies on a larger-scale LLM, as this stage demands stronger reasoning and synthesis capabilities. By orchestrating models of varying capacities in this way, SpecRCA achieves high exploration diversity while simultaneously reducing inference cost and latency.

We conduct preliminary experiments on the AIOps 2022 dataset. In terms of accuracy, SpecRCA surpasses the state-of-the-art methods in failure localization by approximately 12.14\%. In terms of efficiency, SpecRCA is able to generate a complete diagnosis report within 20 seconds, significantly faster than existing approaches. These preliminary results demonstrate that SpecRCA achieves both higher diagnostic accuracy and substantially reduced inference latency, highlighting its potential for real-world deployment.

\section{Methodology}

Figure~\ref{fig: architecture} provides a high-level view of the SpecRCA workflow. Upon detecting an anomaly in a microservice system, the framework collects abnormal operational data and a preceding segment of normal data (metrics, traces, and logs) to establish a behavioral baseline. Both datasets are then processed by the Hypothesis Drafting module, which efficiently generates a broad set of candidate root causes. These hypotheses are passed to the Root Cause Verifier for fine-grained validation, yielding reasoning and supporting evidence. Finally, the Diagnosis Synthesizer aggregates verification results, resolves inconsistencies, and outputs a coherent, interpretable diagnosis report.

\subsection{Hypothesis Drafting}

\begin{figure}[htbp]
	\centering
	\includegraphics[width=1\linewidth]{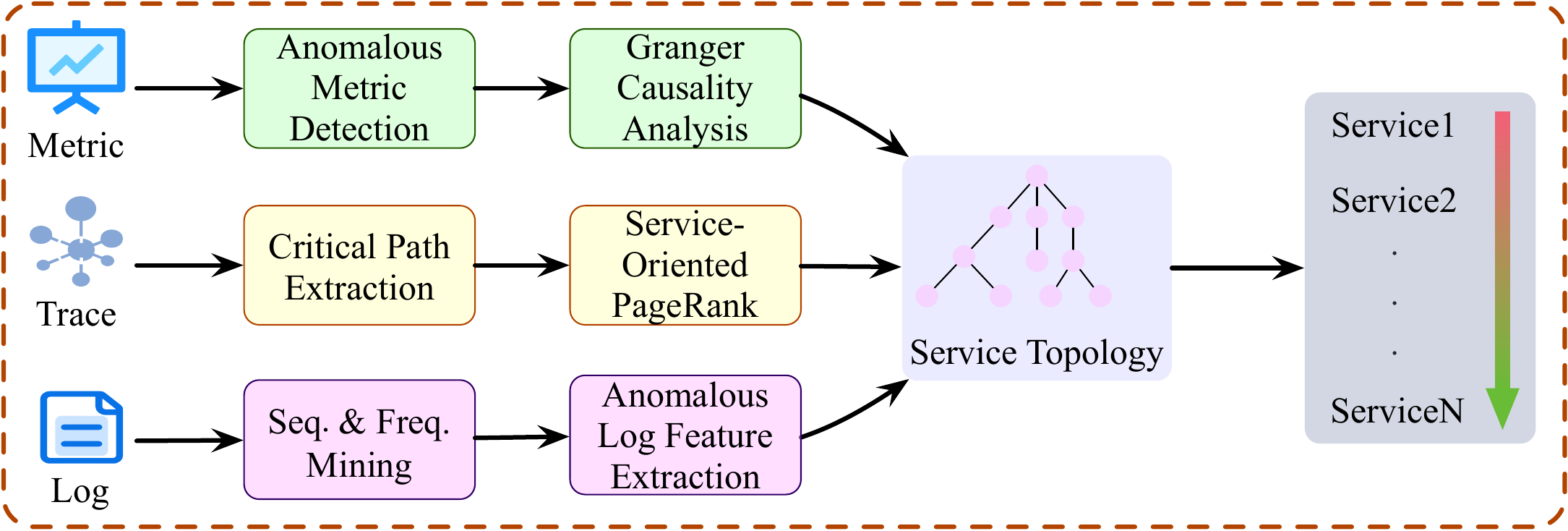}
	\vspace{-1.2em}
	\caption{Hypothesis Drafting}
	\label{fig: hypothesis-drafting}
	\vspace{-1.2em}
\end{figure}

The goal of Hypothesis Drafting is to generate a comprehensive and prioritized set of candidate root causes from heterogeneous system signals. As illustrated in Figure~\ref{fig: hypothesis-drafting}, the module consists of three pipelines (metrics, traces and logs) that produce modality-specific suspiciousness scores which are then aligned to services and fused under topological constraints.

\textbf{Metrics Analysis.} This component consists of two main modules: \textit{Anomalous Metric Detection} and \textit{Granger Causality Analysis}, which together identify suspicious services based on deviations from normal operation and inter-service dependencies.

\textit{Anomalous Metric Detection} quantifies the deviation of service-level metrics during abnormal periods. For each service $i$, the metrics collected in the abnormal period are denoted by $\mathbf{X}_{\text{abn}}^{(i)} = \{x_t^{(i)}\}_{t=1}^{T_\text{abn}}$, and the baseline metrics during normal operation by $\mathbf{X}_{\text{norm}}^{(i)} = \{x_t^{(i)}\}_{t=1}^{T_\text{norm}}$. The service-level anomalousness score is computed via the 1-Wasserstein distance, as calculated in Equation~\ref{eq: amd-score-metric}, where $W_1(P,Q) = \inf_{\gamma \in \Gamma(P,Q)} \mathbb{E}_{(x,y)\sim \gamma} [|x-y|]$.  

\begin{equation} 
	\mathrm{Score}_{\text{metric}}(i) = W_1\big(\mathbf{X}_{\text{norm}}^{(i)}, \mathbf{X}_{\text{abn}}^{(i)}\big)
	\label{eq: amd-score-metric} 
\end{equation}

\textit{Granger Causality Analysis} leverages the temporal relationships between services to refine candidate root causes. For two services $i$ and $j$ with metric time series $x^{(i)}_t$ and $x^{(j)}_t$, service $i$ is considered to Granger-cause service $j$ if it satisfies the condition in Equation~\ref{eq: satisfy-granger-cause}, where $\mathcal{F}_{t-1}$ is the full past information and $\mathcal{F}_{t-1}^{(-i)}$ excludes the history of service $i$.

\begin{equation} 
	\mathrm{Var}\big(x^{(j)}_t \mid \mathcal{F}_{t-1}^{(-i)}\big) > \mathrm{Var}\big(x^{(j)}_t \mid \mathcal{F}_{t-1}\big)
	\label{eq: satisfy-granger-cause} 
\end{equation}

Significant causal links are retained to adjust the anomalousness scores, producing a causality-weighted metric score, as illustrated in Equation~\ref{eq: granger-score}, where $\mathcal{C}(i)$ is the set of downstream services of $i$ in the service topology, $\beta$ is a tunable influence coefficient, and $\mathbf{1}_{\text{Granger}(i\to j)}$ indicates whether the causal link is significant.

\begin{equation} 
	\mathrm{Score}_{\text{metric}}^{\text{causal}}(i) = \mathrm{Score}_{\text{metric}}(i) \cdot \left(1 + \sum_{j \in \mathcal{C}(i)} \beta \cdot \mathbf{1}_{\text{Granger}(i\to j)} \right)
	\label{eq: granger-score} 
\end{equation}

\textbf{Trace Analysis.}  
From distributed traces, we construct a call graph $G=(V,E)$ where $V$ denotes services and $E$ represents observed invocations. We first extract the \textit{critical path} $\pi^*$ with the largest latency contribution. For each node $v$ on the path, define the residual latency as Equation~\ref{eq: trace-critical-path}, where $L_\tau(v)$ is the observed latency and $\bar{L}(v)$ is the baseline mean. 

\begin{equation}
	r_\tau(v)=L_\tau(v)-\bar{L}(v),
	\label{eq: trace-critical-path} 
\end{equation}

To rank suspicious services, we apply a trace-weighted \textit{Service-Oriented PageRank} as Equation~\ref{eq: trace-pagerank}, where $M(\tau_u)$ is the accumulated residual latency on the critical path through node $u$, $\mathcal{N}^-(v)$ denotes incoming neighbors of $v$, and $\alpha$ is the damping factor.

\begin{equation}
	PR(v) = (1-\alpha)\frac{1}{|V|} + \alpha \sum_{u \in \mathcal{N}^-(v)} \frac{PR(u)\cdot M(\tau_u)}{\sum_{w \in \mathcal{N}^-(v)} M(\tau_w)}
	\label{eq: trace-pagerank} 
\end{equation}

\textbf{Log Analysis.} Logs are first parsed into structured templates to unify heterogeneous log formats. Let $\mathcal{L}_s = \{l_1, l_2, \dots, l_n\}$ denote the sequence of templates for service $s$.  

\textit{Sequence Mining.} For each service, we model the normal log sequences $\mathcal{S}_{\text{norm}}(s)$ using sliding-window subsequences of length $k$. During anomalies, we observe sequences $\mathcal{S}_{\text{abn}}(s)$. The sequence anomaly score for a subsequence $q$ is defined as Equation~\ref{eq: log-seq}, where $P_{\text{norm}}(q)$ and $P_{\text{abn}}(q)$ are the probabilities of observing $q$ under normal and abnormal conditions, respectively. This captures missing or newly emerging sequences indicative of anomalous behavior. 

\begin{equation}
	\mathrm{SeqAnom}(q) = \big|\log\!\big(\frac{P_{\text{abn}}(q)+\epsilon}{P_{\text{norm}}(q)+\epsilon}\big)\big|,
	\label{eq: log-seq} 
\end{equation}

\textit{Frequency Mining.} To detect rare or bursty templates, we compute the template anomalousness score as Equation~\ref{eq: log-burst}, where $\mathrm{freq}_{\cdot}(t)$ counts occurrences of template $t$ in the corresponding dataset.  

\begin{equation}
	\mathrm{Burst}(t) = \frac{\mathrm{freq}_{\text{abn}}(t) - \mathrm{freq}_{\text{norm}}(t)}{\sqrt{\mathrm{freq}_{\text{norm}}(t)+\epsilon}},
	\label{eq: log-burst} 
\end{equation}

\textbf{Anomalous Log Feature Extraction.}  
The service-level log anomalousness integrates sequence and frequency evidence as Equation~\ref{eq: log-result}, where $\mathcal{T}(s)$ denotes the set of templates associated with service $s$. This score provides a unified measure of anomalous patterns in both sequences and individual template occurrences.

\begin{equation}
	\mathrm{Anom}_{\text{log}}(s) = \sum_{q \in \mathcal{S}_{\text{abn}}(s)} \mathrm{SeqAnom}(q) + \sum_{t \in \mathcal{T}(s)} \mathrm{Burst}(t),
	\label{eq: log-result} 
\end{equation}

\textbf{Topology-Guided Integration.}  
Candidate evidence from metrics, traces, and logs is fused into a service-level suspiciousness ranking, exploiting the service dependency topology $\mathcal{T} = (S,E)$.  

Each service $s \in S$ receives an initial score, as shown in Equation~\ref{eq: score-init}, where $w_m, w_t, w_l$ are modality weights.  
\begin{equation}
	\mathrm{Score}_0(s) = w_m \cdot \max_j \mathrm{GC}_{j \to s} + w_t \cdot PR(s) + w_l \cdot \mathrm{Anom}_{\text{log}}(s),
	\label{eq: score-init} 
\end{equation}

Suspiciousness propagates over topology to highlight hidden root causes, as calculated as Equation~\ref{eq: score-prop}, with $\tilde{A} = D^{-1}A$ the normalized adjacency matrix and $\alpha$ controlling propagation strength.  

\begin{equation}
	\mathbf{Score}^{(k+1)} = \alpha \cdot \tilde{A}^\top \mathbf{Score}^{(k)} + (1-\alpha)\cdot \mathbf{Score}_0
	\label{eq: score-prop} 
\end{equation}

Finally, to capture shared downstream dependencies, a common-child regularizer adjusts scores as Equation~\ref{eq: score-final}, where $\mathcal{C}(s)$ is the set of direct downstream children of $s$, and $\lambda$ balances the influence.

\begin{equation}
	\mathrm{Score}(s_i) \leftarrow \mathrm{Score}(s_i) + \lambda \sum_{j \neq i} \frac{|\mathcal{C}(s_i)\cap\mathcal{C}(s_j)|}{|\mathcal{C}(s_i)\cup\mathcal{C}(s_j)|} \cdot \mathrm{Score}(s_j),
	\label{eq: score-final} 
\end{equation}

\subsection{Root Cause Verifier}

\begin{figure}[htbp]
	\centering
	\includegraphics[width=1\linewidth]{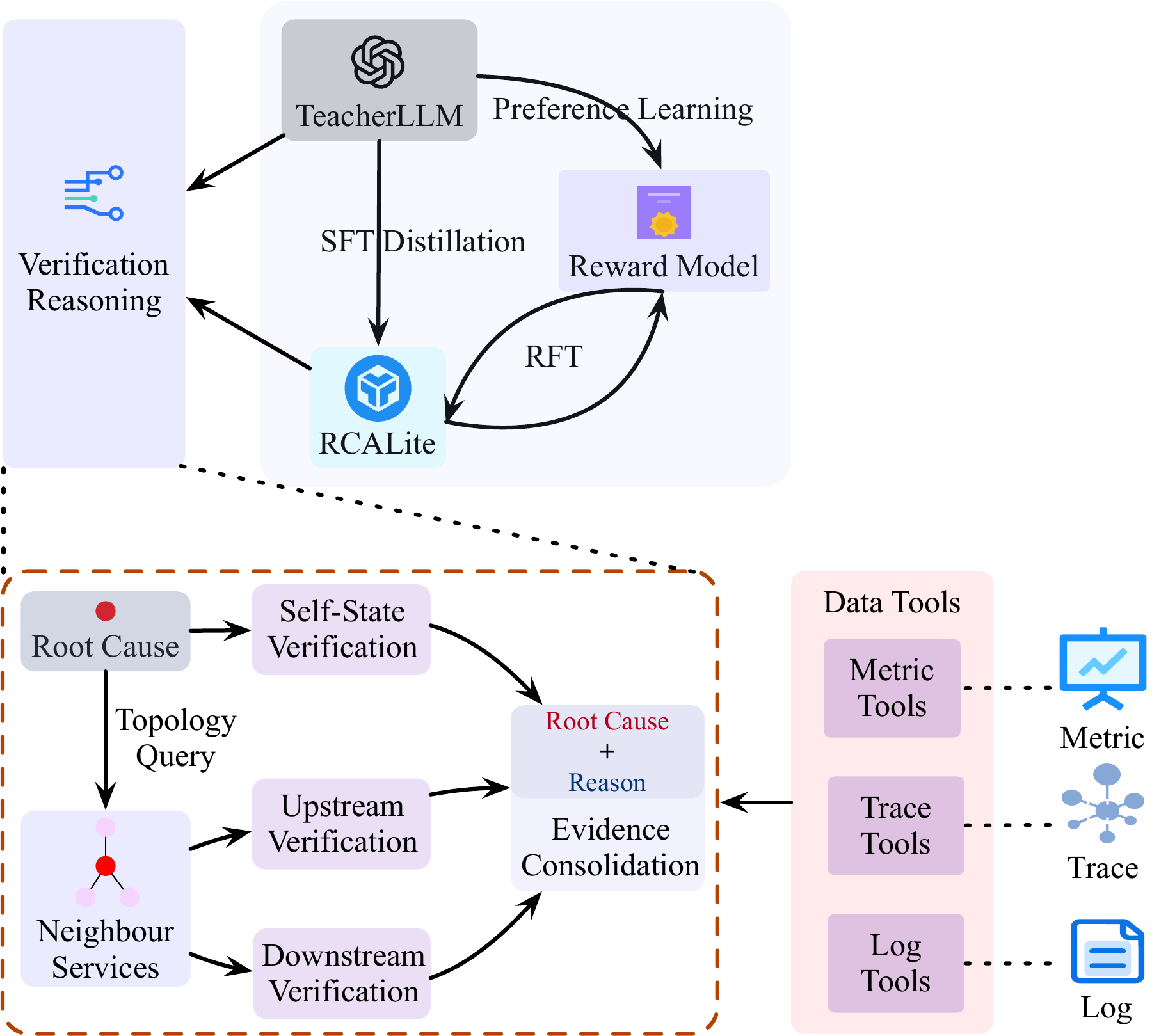}
	\vspace{-1.2em}
	\caption{Root Cause Verifier}
	\label{fig: path-diagnosis}
	\vspace{-2.0em}
\end{figure}

The Root Cause Verifier is centered on the \textbf{RCALite} model, which performs verification reasoning on candidate root cause hypotheses using heterogeneous system data. As shown in Figure~\ref{fig: path-diagnosis}, the overall workflow consists of three stages: distillation from a Teacher LLM, reward model learning, and reinforcement fine-tuning.

\vspace{-1.0em}

\paragraph{Teacher LLM Distillation.}  
A high-capacity Teacher LLM, $\mathcal{M}_T$ (Claude-3.5-Sonnet in this work), generates structured verification reasoning for each candidate hypothesis $h \in \mathcal{H}$. RCALite, denoted $\mathcal{M}_R$, is trained via supervised fine-tuning to mimic $\mathcal{M}_T$, as shown in Equation~\ref{eq: distillation}, where $\ell(\cdot,\cdot)$ is a token-level cross-entropy loss.

\begin{equation}
	\mathcal{L}_{\text{SFT}} = \sum_{h \in \mathcal{H}} \ell\Big(\mathcal{M}_R(h), \mathcal{M}_T(h)\Big)
	\label{eq: distillation} 
\end{equation}

\paragraph{Reward Model and Reinforcement Fine-Tuning.}  
To enhance reasoning reliability, a reward model $\mathcal{R}_\theta$ is learned via preference data from $\mathcal{M}_T$, and RCALite is fine-tuned using reinforcement learning. The training objective is to maximize expected reward, as illustrated in Equation~\ref{eq: grpo}, where $\pi_\phi$ denotes RCALite's policy for generating verification reasoning traces, optimized using GRPO.

\begin{equation}
	\mathcal{J}(\phi) = \mathbb{E}_{\mathcal{R}_R \sim \pi_\phi(\cdot|h)} \big[ \mathcal{R}_\theta(\mathcal{R}_R) \big],
	\label{eq: grpo} 
\end{equation}

\paragraph{Verification Reasoning.}  

During inference, RCALite receives a candidate root cause hypothesis $h$ as input and performs structured verification reasoning to assess its validity. The verification process consists of four main components:

\begin{enumerate}[leftmargin=*]
	\item \textit{Self-State Verification:} RCALite evaluates the internal consistency of the hypothesis with respect to the observed abnormal and normal system data, examining whether the hypothesis alone can account for the detected anomalies.
	
	\item \textit{Upstream Verification:} For each parent service $p$ of the candidate service in the service topology, RCALite investigates whether anomalies in $p$ could explain the observed system behavior, thus validating potential upstream influences.
	
	\item \textit{Downstream Verification:} For each child service $c$, RCALite analyzes whether the observed anomalies in $c$ suggest that the true root cause might reside downstream rather than at the candidate service itself.
	
	\item \textit{Evidence Consolidation:} RCALite integrates the results from self-state, upstream, and downstream verification to produce a diagnosis. Based on the gathered evidence, it determines which component within the local neighborhood of the candidate hypothesis is most likely the actual root cause, providing detailed reasoning and supporting observations for its conclusion.
\end{enumerate}

\section{Preliminary Evaluation}

To evaluate SpecRCA, we assess its feasibility and effectiveness on the AIOPS 2022 dataset. For this preliminary evaluation, RCALite is distilled from Claude-3.5 using Llama3.2-3B as the base model. At this stage, we have not yet performed the subsequent RFT process. The modality weights are set as $w_m = 0.3$, $w_t = 0.4$, and $w_l = 0.2$. For comparison, we evaluate SpecRCA against RCAgent and mABC, both of which are based on Qwen-2.5-Plus.

\vspace{-0.8em}
\begin{table}[htb]
	\setlength{\tabcolsep}{3.5pt}
	\centering
	\caption{Evaluation Results. Accuracy measured by Recall@K and MRR; speed measured in seconds per query (s/q).}
	\label{tab: evaluation}
	\vspace{-1.2em}
	\begin{tabular}{c|cccc|c}
		\toprule
		Approach & Recall@1 & Recall@3 & Recall@5 & MRR & Speed (s/q) \\
		\midrule
		RCAgent & 22.10 & 28.40 & 30.25 & 23.95 & 52.79 \\
		mABC & 34.19 & 42.13 & 44.51 & 38.46 & 83.17 \\
		SpecRCA & \textbf{61.34} & \textbf{75.72} & \textbf{81.63} & \textbf{62.64} & \textbf{9.89} \\
		\bottomrule
	\end{tabular}
	\vspace{-0.8em}
\end{table}

The evaluation results are summarized in Table~\ref{tab: evaluation}. SpecRCA significantly outperforms both RCAgent and mABC across all accuracy metrics, achieving a Recall@1 of 61.34\%, Recall@3 of 75.72\%, Recall@5 of 81.63\%, and an MRR of 62.64\%. In terms of efficiency, SpecRCA demonstrates a substantial advantage, generating a diagnosis in only 9.89 seconds per query, which is considerably faster than the baseline approaches. These preliminary results highlight the effectiveness of SpecRCA in both accuracy and inference speed.

\section{Conclusion}

To address the key limitations of current LLM-based RCA methods—(a) limited exploration diversity and (b) overreliance on large-scale LLMs—we propose SpecRCA, a speculative root cause analysis framework for microservices that leverages pathwise parallelism and follows a hypothesize-then-verify paradigm. The effectiveness of this approach is demonstrated through preliminary evaluation on a prototype implementation.

\section{Future Plans}

We plan to pursue two main directions. First, we aim to further refine the implementation of SpecRCA, including applying RFT to fine-tune RCALite, which has so far only undergone SFT distillation, and conducting additional experiments across more datasets to comprehensively evaluate SpecRCA’s performance. Second, the Diagnosis Synthesizer currently represents the primary bottleneck in inference speed. To address this, we plan to leverage diffusion-based large language models, which can accelerate generation through parallel decoding and thus significantly improve throughput.

\begin{acks}
	This work is supported by Key RD Project of Guangdong Province, China (No.2020B010164003).
\end{acks}
\bibliographystyle{ACM-Reference-Format}
\balance
\bibliography{sample-base}

\end{document}